\documentclass[aps,pre,twocolumn,
showpacs,
preprintnumbers,amsmath,amssymb,
epsf,superscriptaddress,groupedaddress,tightenlines,nofootinbib]{revtex4}
\usepackage{epsfig,graphicx}

\usepackage{graphicx}
\usepackage{amsmath}
\usepackage{bm}

\newcommand{\beqa}{\begin{eqnarray}}
\newcommand{\eeqa}{\end{eqnarray}}
\newcommand{\beq}{\begin{equation}}
\newcommand{\eeq}{\end{equation}}
\newcommand{\bsp}{\begin{split}}
\newcommand{\esp}{\end{split}}
\newcommand{\bal}{\begin{align}}
\newcommand{\eal}{\end{align}}

\newcommand{\E}{\mathbb{E}}

\begin{document}

\allowdisplaybreaks
\vspace*{10pt}
\title{Equivalence of interest rate models and lattice gases}

\author{Dan Pirjol}
\affiliation{J.~P.~Morgan, 277 Park Avenue, New York, NY 10172}
%
\begin{abstract} 
\noindent
We consider the class of short rate interest rate models for which the short rate
is proportional to the exponential of a Gaussian Markov 
process $x(t)$ in the terminal measure $r(t) = a(t)\exp(x(t))$.
These models include the Black, Derman, Toy 
and Black, Karasinski models in the terminal measure. We show that such 
interest rate models are equivalent with lattice gases with attractive 
two-body interaction $V(t_1,t_2)= -\mbox{Cov }(x(t_1),x(t_2))$.   
We consider in some
detail the Black, Karasinski model with $x(t)$ an Ornstein, Uhlenbeck
process, and show that it is similar with a lattice gas model considered
by Kac and Helfand, with attractive long-range 
two-body interactions $V(x,y) = -\alpha 
(e^{-\gamma |x - y|} - e^{-\gamma (x + y)})$.
An explicit solution for the model is given as a sum over the states
of the lattice gas, which is used to show that the model has a 
phase transition similar to that found previously in the Black, Derman, Toy 
model in the terminal measure.
\end{abstract}
\maketitle

\section{Introduction}

We consider in this paper the class of one-factor interest 
rate models with log-normally distributed short rate in the terminal measure. 
In these models the short rate is driven by one Gaussian Markov process $x(t)$.
Such a process is defined by two conditions: i) for any set of times
$t_1<t_2< \cdots < t_k$, the values $(x(t_1),x(t_2),\cdots , x(t_k))$ have
a joint normal distribution; ii) the evolution of $x(s)$ for all $s>t$
depends only on $x(t)$. It can be shown that the most general
process of this type is a time-changed Brownian motion, and includes
the Ornstein-Uhlenbeck process as a particular case \cite{Doob}.

This class of models includes the Black, Derman, Toy (BDT) \cite{BDT} model, 
and the Black-Karasinski (BK) \cite{BK} model, formulated in the terminal measure.
The terminal measure is sometimes used in practice for these models \cite{AP},
as opposed to the spot measure in which the models were originally formulated, 
due to the ease of calibration and simulation. 
Such models have been also proposed as approximations to the Libor market model
\cite{LMM,LMM1}, and as particular parametric realizations of Markov functional 
models \cite{1,AP}. 
A choice of measure amounts to a distributional assumption for the 
dynamical variables of the model. See \cite{BR} for a readable introduction to 
the related concepts of martingales and measure for stochastic processes,
and their relation to arbitrage pricing theory.

In this paper we show that these interest rate models are equivalent with 
lattice gases with attractive
two-body interaction $V(t_1,t_2)=-\mbox{Cov}(x(t_1),x(t_2))$, placed in
an external potential. The solution of the models can be expressed
explicitly as an expression for the one-step zero coupon bond given by
a sum over occupation numbers in the lattice gas. The expectation 
values required for the simulation of the model correspond to thermodynamical
potentials in the lattice gas model. 

We discuss in some detail the Black, Karasinski model with constant mean
reversion $\gamma$, which is equivalent to a lattice gas with attractive two-body
interaction $V(x,y) = -\alpha (e^{-\gamma |x - y|} - e^{-\gamma (x + y)})$.
This is similar to a lattice gas model considered by Kac \cite{Kac}, 
Kac, Uhlenbeck, Hemmer \cite{KUH} and Kac, Helfand \cite{KacHelfand,Helfand}. 
This  model generalizes the BDT model in the terminal measure, which
corresponds to $\gamma=0$, and is equivalent with a Coulomb lattice
gas with attractive two-body interactions. The latter model was studied
in Ref.~\cite{PhaseTransition}, where it was shown  that it displays 
discontinuous behaviour in volatility, which is similar to a phase transition
in condensed matter physics \cite{EHS,NG}.

The equivalence with the lattice gas models suggests alternative
simulation methods for these interest rate models, which express
expectation values as sums over the states of the lattice gas.
For small lattices this can be done by explicit summation over the
lattice gas states, while for bigger lattices efficient numerical methods are
available from statistical mechanics, such as Gibbs sampling and the 
Metropolis algorithm. We illustrate this approach by a numerical study
of the BK model, which shows that the volatility phase transition observed in the
BDT model in Ref.~\cite{PhaseTransition} persists also for this model.

\section{The interest rate model}

We consider a short rate interest rate model in discrete time. 
The model is defined on a finite set of dates 
\begin{eqnarray}
0 = t_0 < t_1 < \cdots < t_n
\end{eqnarray}
For simplicity we will assume that $t_i$ are equally spaced,
and denote $\tau = t_{i+1} - t_i$ with $i=0,1,\cdots , n-1$.

The fundamental dynamical quantities of the model are the zero coupon bonds 
$P_{i,j}\equiv P_{t_i,t_j}$. They are defined as the price at time $t_i$ of
a payment of 1 made at time $t_j$. They are stochastic quantities, 
and can be expressed as functions of an one-dimensional Markov 
process $x(t)$. For definiteness we consider in the following that $x(t)$ is an
Ornstein-Uhlenbeck process with zero mean reversion level
\begin{eqnarray}\label{OU}
dx(t) = -\gamma x(t) dt + \sigma dW(t)\,.
\end{eqnarray}
The mean and variance of $x(t)$ conditional on $x(0)=0$ are
\begin{eqnarray}
&& \mathbb{E}[x(t)|x(0)=0]=0 \\
\label{x2}
&& \mathbb{E}[x^2(t)|x(0)=0] = \frac{\sigma^2}{2\gamma}(1 - e^{-2\gamma t}) 
\equiv G(t)\,. 
\end{eqnarray}
The arguments of this paper can be easily 
extended to the more general case of $x(t)$ an arbitrary Gaussian Markov 
process. By the Doob's representation, the
most general Gaussian Markov process can be represented as a time-modified
Brownian motion \cite{Doob}
\begin{eqnarray}
x(t) = f(t) \int_0^t g(s) dW(s)
\end{eqnarray}
with $f(t),g(t)$ deterministic functions of time, and $W(t)$ a Brownian motion. 

We define the Libor rate (or simply Libor) for the $(t_i,t_{i+1})$ period as
\begin{eqnarray}
L_{i} = \tau^{-1} \Big( \frac{1}{P_{i,i+1}} - 1 \Big)\,.
\end{eqnarray}
The model is defined by specifying the functional dependence of the
Libor rate $L_i$ on the Markov driver $x(t_i)$
\begin{eqnarray}\label{Lipdf}
L_i = \tilde L_i \exp\Big( x(t_i) - \frac12 G(t_i) \Big)
\end{eqnarray}
where $\tilde L_i$ are constants to be chosen such that the initial
yield curve $P_{0,t}$ is correctly reproduced. This implies that
the Libors $L_i$ are log-normally distributed in the terminal measure.

This model is similar with the Black-Karasinski model \cite{BDT,BK},
up to the difference that the latter is usually formulated in the
risk-neutral measure, while in the model considered here the short rate
$L_i$ is expressed in terms of $x(t)$ defined in the terminal measure.

In the limit when the time step is taken to zero $\tau \to 0$,
this model becomes a continuous time short rate model, and the short rate 
$r(t) = \lim_{\tau\to 0} L_{t/\tau}(t)$ satisfies the stochastic differential
equation
\begin{eqnarray}
\frac{dr(t)}{r(t)} = (a(t) - \gamma \ln r(t)) dt + \sigma dW(t)
\end{eqnarray}
with $a(t)$ a function depending on $\tilde L_i$ and $\sigma$. 
We recognize this as the short rate evolution in the Black-Karasinski 
model \cite{BK}.

\subsection{Explicit solution of the model}

According to the fundamental theorem of arbitrage pricing theory \cite{BR},
the price of a financial asset $V(t)$ expressed in units of a simpler 
asset $N(t)$  (called numeraire) is a martingale. The mathematical
statement of this result is expressed as 
\begin{eqnarray}\label{mart}
V(t)/N(t) = \mathbb{E}[V(T)/N(T)|{\cal F}_t]\,,
\end{eqnarray}
for any $t<T$.
This holds under fairly general assumptions, among which market completeness
is the most important one. Speaking loosely this means that the model contains 
sufficiently many tradeable instruments to allow any possible payout to be 
reproduced as a combination thereof.

The choice of the numeraire $N(t)$ is not unique, and any particular choice
defines a measure for the stochastic process followed by the discounted
asset prices $V(t)/N(t)$. Two particular choices are most common in the 
context discussed here.
The spot measure, or the risk-neutral measure, takes $N(t)$ to be the money market 
account at time $t$, while the terminal measure (or $t_n$-forward measure)
takes $N(t)=P_{t,n}$  to be the zero coupon bond maturing at time $t_n$.
Once the condition (\ref{Lipdf}) is imposed, different measure choices  
produce different observable distributional properties of the dynamical 
quantities of the model (rates and bonds), and thus effectively correspond to
different models. 

We will work in the terminal measure in the following.
It is convenient to introduce the zero coupon bond prices divided 
by the numeraire $P_{t,n}$, which will be denoted as
$\hat P_{i,j} = P_{i,j}/P_{i,n}$.
They are martingales in the terminal measure, and thus satisfy
the condition (\ref{mart}), which reads explicitly
\begin{eqnarray}\label{martingale}
\hat P_{i,j} = \E\Big[\frac{P_{k,j}}{P_{k,n}} | {\cal F}_i\Big]
\end{eqnarray}
for all $i < k < j \leq n$. The one-step discounted zero bond $\hat P_{i,i+1}(x_i)$
will play an important role in writing the analytical solution of this model.
It satisfies a few conditions, following from the martingale condition
(\ref{martingale}). First, its expectation value is known in terms of the
initial yield curve
\begin{eqnarray}\label{exp1}
\E[\hat P_{i,i+1}(x_i)] = \hat P_{0,i+1} \,.
\end{eqnarray}

It also satisfies the two conditions
\begin{eqnarray}\label{mart1}
&& \hat P_{i,i+1}(x_i) = \E[ \hat P_{i+1,i+2}(x_{i+1})
(1 + L_{i+1}(x_{i+1})\tau) | {\cal F}_i] \nonumber \\
&& \\
\label{mart2}
&& \hat P_{0,i} = \E[\hat P_{i,i+1}(x_i) (1 + L_i(x_i)\tau) ]
\end{eqnarray}
The first condition (\ref{mart1}) determines recursively the functional
form of $\hat P_{i,i+1}(x_i)$, starting with $\hat P_{n-1,n}=1$ and proceeding
backwards in time. This is given explicitly as a conditional expectation
value
\begin{eqnarray}\label{PiiProd}
\hat P_{i,i+1}(x_i) &=& \mathbb{E}\Big[ \prod_{k=i+1}^{n-1} (1 + \tilde L_k\tau
e^{x_k - \frac12 G_k}) | {\cal F}_i\Big] \,.
\end{eqnarray}

The second condition (\ref{mart2}) can be used to determine 
$\tilde L_i$ also recursively, once $\hat P_{i,i+1}(x_i)$ has been determined,
using the relation
\begin{eqnarray}\label{Litilde}
\tilde L_i = \frac{\hat P_{0,i} - \hat P_{0,i+1}}
{\E[\hat P_{i,i+1}\exp(x_i - \frac12 G_i)] \tau} \,.
\end{eqnarray}
For simplicity we denote the value of the Markov driver at time $t_i$ 
as $x_i \equiv x(t_i)$, and its variance as $G(t_i) = G_i$.

We will state in the following the closed form of the solution of this model.
The solution expresses the discounted one-step 
zero coupon bonds $\hat P_{i,i+1}(x_i)$ as a sum of terms containing
$0,1,2,\cdots , n-i-1$ $\tilde L_j$ factors. Writing the first few terms 
explicitly this is given by 
\begin{widetext}
\begin{eqnarray}\label{Pii}
&& \hat P_{i,i+1}(x_i) = 1 + \sum_{j=i+1}^{n-1}
\tilde L_j \tau \exp( w^{j-i} x_i - \frac12 w^{2(j-i)} G_i ) \nonumber \\
&& + \sum_{j>k=i+1}^{n-1}\tilde L_j\tilde L_k \tau^2
\exp\Big( (w^{j-i} + w^{k-i}) x_i 
- \frac12 (w^{j-i} + w^{k-i})^2 G_i + X_{jk}\Big) + \cdots \\
&& +  \sum_{k\leq n-i-1} \sum_{S_k \in T_i}
\tilde L_{j_1}\tilde L_{j_2}\cdots \tilde L_{j_k} \tau^k
\exp\Big( \sum_{a=1}^k w^{j_a-i}  x_i -
\frac12 (\sum_{a=1}^k w^{j_a-i})^2 G_i + 
\sum_{1<a<b<k} X_{j_a, j_b} \Big) \nonumber \,.
\end{eqnarray}
\end{widetext}
We denoted here the weight $w=\exp(-\gamma\tau)$, and the auto-covariance
of the Markov process $x(t)$ as
\begin{eqnarray}\label{covx}
X_{jk} &=& \mbox{Cov}(x(t_j), x(t_k))\\
 &=& \frac{\sigma^2}{2\gamma} 
(e^{-\gamma |t_j - t_k|} - e^{-\gamma (t_j + t_k)})\,.\nonumber
\end{eqnarray}
The general term in Eq.~(\ref{Pii}) containing $k\leq n-i-1$ factors of 
$\tilde L_j$ is given by a sum over all subsets $S_k = \{ j_1, j_2, \cdots , j_k\}$
of $k$ indices chosen from the $n-i-1$ indices 
$T_i \equiv \{i+1,i+2, \cdots , n-1\}$.

In the limit of zero mean reversion $\gamma \to 0$, we have $w=1$ 
and $G(t) = \sigma^2 t$, and the expression (\ref{Pii}) simplifies drastically. 
In this limit all terms with the same number
of $\tilde L_j$ factors have the same functional dependence of $x_i$,
and we recover the simple form obtained in Ref.~\cite{PhaseTransition}
\begin{eqnarray}
\hat P_{i,i+1}(x) = \sum_{j=0}^{n-1} c_j^{(i)} e^{j x_i -\frac12 j^2 G_i}
\end{eqnarray}
where the coefficients $c_j^{(i)}$ are given by
\begin{eqnarray}\label{ciresult}
c_k^{(i)} = \sum_{S_k}
\tilde L_{j_1}\tilde L_{j_1}\cdots \tilde L_{j_k} \tau^k
\exp( \sum_{1<a<b<k} X_{j_a, j_b} )
\end{eqnarray}
where $X_{j,k} = \sigma^2 \mbox{min}(t_j, t_k)$. In \cite{PhaseTransition}
these coefficients were determined recursively from a recursion relation,
see Eq.~(12) in Ref.~\cite{PhaseTransition}.
Equation (\ref{ciresult}) gives an explicit solution of this recursion relation.

An important role is played in this model by the expectation values of the
form
\begin{eqnarray}\label{Niphidef}
&& N_i(\phi) = \E[\hat P_{i,i+1} e^{\phi x_i - \frac12 \phi^2  G_i}]\\
&& = 
1 + \sum_{j=i+1}^{n-1}
\tilde L_j \tau \exp(\phi w^{j-i} G_i) + \cdots \nonumber\\
&& \qquad + \sum_{k\leq n-i-1}\sum_{S_k}
\tilde L_{j_1}\tilde L_{j_2}\cdots \tilde L_{j_k} \tau^k \nonumber \\
&& \times \exp( \phi G_i \sum_{a=1}^k w^{j_a-i} + \sum_{1<a<b<k} X_{j_a, j_b} ) 
\nonumber \,.
\end{eqnarray}
We enumerate in the following the applications of these expectation values with 
$\phi = 0,1,\cdots$. 

The expectation value of $\hat  P_{i,i+1}$ (corresponding to $\phi=0$)
is constrained by the requirement
that the initial yield curve $P_{0,i}$ is correctly reproduced, see (\ref{exp1}).
\begin{eqnarray}\label{N0}
&& \E[\hat P_{i,i+1}] = \hat P_{0,i+1} = 
1 + \sum_{j=i+1}^{n-1} \tilde L_j \tau + \cdots \\
&& + \sum_{S_k}
\tilde L_{j_1}\tilde L_{j_1}\cdots \tilde L_{j_k} \tau^k
\exp( \sum_{1<a<b<k} X_{j_a, j_b} ) + \cdots \nonumber
\end{eqnarray}
The sum on the right-hand side is linear in $\tilde L_{i+1}$ and thus
can be used to solve explicitly for this constant, provided that all
$\tilde L_j$ with $j=i+2, \cdots, n-1$ are already known. This is given  
in Eq.~(\ref{Litilde}) in a form more convenient for practical calculation.

The $\phi=1$ expectation value appears in the calculation of the 
convexity-adjusted
Libors $\tilde L_i$ Eq.~(\ref{Litilde}), which can be written
equivalently as
\begin{eqnarray}\label{tildeLi}
\tilde L_i = \hat P_{0,i+1} L_i^{\rm fwd} \frac{1}{N_i(1)}\,.
\end{eqnarray}
Finally, $N_i(j)$ with $j\in \mathbb{Z}_+, j>1$ determines the 
$j-$th moment of the Libor distribution in its natural (forward)
measure according to the relation \cite{2}
\begin{eqnarray}
\mathbb{E}_{i+1}[(L_i)^j] &=& \frac{1}{\hat P_{0,i+1}} (\tilde L_i)^j
\mathbb{E}_n [ \hat P_{i,i+1} e^{j x_i - \frac12 j G_i} ] \nonumber\\
&=&
\frac{1}{\hat P_{0,i+1}} (\tilde L_i)^j e^{-\frac12 (j-j^2) G_i} N_i(j)
\end{eqnarray}

In the limit of zero mean-reversion $\gamma\to 0$ the above
expectation values are given by simple expressions \cite{PhaseTransition}
\begin{eqnarray}
N_i(\phi) = \E[\hat P_{i,i+1} e^{\phi x_i -\frac12\phi^2 G_i}] =  
\sum_{j=0}^{n-i-1} c_j^{(i)} e^{j \phi^2 \sigma^2 t_i}\,.
\end{eqnarray}

For  sufficiently small volatility $\sigma$, the expectation values 
$N_i(\phi)$ given in Eq.~(\ref{Niphidef}) can be computed in an
expansion of the small parameter $\tilde L_i\tau \ll 1$, and keeping
only the terms linear in this parameter is sufficient for most
applications. In this approximation we have
\begin{eqnarray}
N_i(\phi)  = 1 + \sum_{j=i+1}^{n-1} L_j^{\rm fwd} \tau e^{\phi w^{j-i} G_i}
+ O((L_k^{\rm fwd}\tau)^2)
\end{eqnarray}
The distribution of the Libors in their natural measure is 
approximatively log-normal and the ATM caplet volatility is 
\begin{eqnarray}
\sigma_{\rm LN}^2 = \frac{G(t_i)}{t_i}\,.
\end{eqnarray}

In the model with zero mean reversion $\gamma=0$, it
was noted in Ref.~\cite{PhaseTransition} that for volatility $\sigma$
above some critical value, the higher order terms in the
expansion (\ref{Niphidef}) become comparable to  the linear terms
of $O(\tilde L_i \tau)$. 
The actual expansion parameter becomes $\tilde L_i \tau \exp(\sigma^2 t_i)$
and terms of all orders in $L_i^{\rm fwd} \tau$ become important. 
This leads to a discontinuity in the first derivative of the expectation 
value $N_i(\phi)$ with respect to the volatility $\sigma$, which is 
similar to a phase transition in condensed matter physics \cite{EHS,NG}.

In the next section we express the expectation values (\ref{Niphidef})
as averages over the grand canonical ensemble in an equivalent lattice gas model.
This is used to show the existence of a phase 
transition also in this model, using a numerical simulation.

\subsection{Proof}

The result (\ref{Pii}) can be proven using the following 
basic identity. For any numbers
$n_k=0,1$ associated with the ordered sequence of times 
$t \equiv t_0 \leq t_1 \leq t_2 \cdots < t_N$, the following expectation
value with $x(t)$ the Ornstein-Uhlenbeck process (\ref{OU}) is given by
\begin{eqnarray}\label{basic}
&& \mathbb{E}\Big[\exp\Big(\sum_{k=1}^N n_k (x_k -\frac12 
G_k ) \Big) | {\cal F}_t\Big] \\
&& = \exp\Big(
x_t\sum_{k=1}^N n_k e^{-\gamma t_k} 
-\frac12 G_t (\sum_{k=1}^N n_k e^{-\gamma t_k})^2 \nonumber \\
&& \qquad + \frac12 \sum_{j\neq k=1}^N X_{j,k} n_j n_k
\Big) \nonumber
\end{eqnarray}
where $X_{j,k}$ is the covariance of the process $x(t)$ given above in
Eq.~(\ref{covx}).
This is a slight generalization of an identity used in 
Ref.~\cite{Kac,KacHelfand} to compute the partition function of a lattice gas  
with exponential interaction. It can be easily generalized to the case of a
general Gaussian Markov process $x(t)$.

The discounted one-step bond $\hat P_{i,i+1}(x_i)$ is given by the
conditional expectation (\ref{PiiProd}).
Expanding out the product yields
terms with $0, 1, 2, \cdots$ factors of $\tilde L_k \tau$, up
to $n-i-1$ factors. There are $\binom{n-i-1}{N}$ terms
containing $N$ such
factors, and they are given by a sum over all subsets $\{ n_k \} = 
\{ n_{k_1}, n_{k_2}, \cdots , n_{k_N}\}$ of
$N$ indices out of the total of $n-i-1$ indices. A generic term has the
form
\begin{eqnarray}
&& \sum_{\{ n_k \}} 
\Pi_{j=1}^{N} (\tilde L_{k_j} \tau) 
 \mathbb{E}[\exp\Big( \sum_{j=1}^{N} n_{k_j} (x_{k_j} - \frac12 G_{k_j} )\Big)|
{\cal F}_i] \nonumber \\
&&  = \sum_{\{ n_k \}} 
\Pi_{j=1}^{N} (\tilde L_{k_j} \tau) \\
&& \times \exp\Big(
x_i\sum_{k=1}^N e^{-\gamma t_{k_j}} n_{k_j} -\frac12 G_i (\sum_{k=1}^N n_{k_j} e^{-\gamma t_{k_j}})^2\nonumber\\
&& \qquad +
\sum_{k_j < k_l} X_{k_j,k_l} n_{k_j} n_{k_l} \Big)\nonumber
\end{eqnarray}
where the expectation value was computed using the identity (\ref{basic}).
This reproduces the terms containing $N$ factors of $\tilde L_k\tau$
in Eq.~(\ref{Pii}). This completes the proof of (\ref{Pii}). 

\section{The lattice gas model}

The interest rate model considered in the previous section is equivalent
with a one-dimensional lattice gas with attractive long-range potential
\begin{eqnarray}\label{Vdef}
V(x,y) = -\alpha
(e^{-\gamma |x - y|} - e^{-\gamma (x + y)})
\end{eqnarray}
The particles of the lattice gas are constrained to sit at positions
$x_i = \tau i$, with $i=1,2,\cdots, n-1$. 
The $n$ sites of the lattice gas are labeled as $j = 0, 1, \cdots, n-1$. 
The sites $j$ are in one-to-one correspondence with the discrete set of 
simulation times $\{ t_j \}$ of the 
interest rate model.
At each site at most one particle can be present. We define $n_j$ the 
occupation number of the site $j$. It can take values 0 or 1,
depending on whether the site $j$ is vacant or occupied. 

The Hamiltonian of the lattice gas model is
\begin{eqnarray}\label{Ham}
H = \sum_{j=1}^{n-1} \varepsilon_j n_j + 
\sum_{j>k=1}^{n-1} \varepsilon_{jk} n_j n_k
\end{eqnarray}
The two-body interaction is
\begin{eqnarray}\label{2body}
\varepsilon_{jk} = 
-\alpha
(e^{-\gamma\tau |j-k|} - e^{-\gamma\tau (j + k)})
\end{eqnarray}
and the single-site energies are
\begin{eqnarray}
\varepsilon_j = - \beta^{-1} \ln(\tilde L_j \tau)\,.
\end{eqnarray}

For the application to the interest rate model we are interested not only 
in the entire lattice system, but also in the subsystem $\mathbb{T}_i$  of the
lattice consisting of the sites $\mathbb{T}_i: \{i+1, \cdots, n-1\}$, in total 
$n_f=n-i-1$ sites. 

Assume that the subsystem $\mathbb{T}_i$ of the lattice gas is placed in a 
position-dependent chemical potential 
\begin{eqnarray}\label{chempot}
\mu^{(i)}(t) = \mu G_i e^{-\gamma (t-t_i)}
\end{eqnarray}

The grand partition function of the subsystem $\mathbb{T}_i$ of the 
lattice gas with the Hamiltonian (\ref{Ham}) and placed in the chemical
potential (\ref{chempot}) is given by
\begin{eqnarray}
{\cal Z}_i(\mu,T) = \sum_{N=0}^{n-i-1} \sum_{S_N} 
\exp\Big(-\beta H + \beta \sum_{j\in S_N}\mu^{(i)}(t_j)\Big)
\end{eqnarray}
The sum over the number of particles $N$ runs from 0 to $n-i-1$, the number
of lattice sites in the subsystem $\mathbb{T}_i$. For each $N$ the sum runs 
over all configurations $S_N$ of $N$ occupied sites, which are subsets of $N$ 
sites of the $n-i-1$ sites in the system $\mathbb{T}_i$.

The correspondence of this lattice gas model
with the interest rate model is realized through the following relation 
between the grand partition function ${\cal Z}_i(\mu,T)$ and 
the expectation value (\ref{Niphidef})
\begin{eqnarray}
N_i(\phi) = {\cal Z}_i(\mu,T) \,,
\end{eqnarray}
provided that the parameters of the lattice gas are related to those of the
interest rate model as
\begin{eqnarray}
\alpha \beta = \frac{\sigma^2}{2\gamma}\,,\qquad
\phi = \beta\mu
\end{eqnarray}

This system is similar to the one-dimensional gas  considered by
Kac \cite{Kac} and by Kac, Uhlenbeck, Hemmer \cite{KUH}. 
A lattice version of the gas model, very similar to that considered
here, was examined by Kac and Helfand in Ref.~\cite{KacHelfand,Helfand}.
More precisely, the latter papers consider a lattice gas, where the
particles occupy a lattice with $N$ nodes and lattice spacing 1, and 
interact by two-body attractive potentials 
$V(|x-y|) =  -\alpha\gamma e^{-\gamma |x-y|}$.
This model has a phase transition in the so-called van der Waals limit,
which is obtained by first going to the thermodynamical limit of large $N$,
followed by the infinite range limit $\gamma \to 0$.
In the van der Waals limit the lattice gas model has a liquid-gas 
phase transition with critical temperature $\beta_c\alpha = \frac12$, 
and the equation of state is given by the van der Waals 
equation supplemented by the equal area rule \cite{KUH}. 

At this point it may be useful to recall a few well-known facts about
phase transitions in one-dimensional systems \cite{Mattis}. Although a phase
transition does not exist in a one-dimensional system with short range
interactions \cite{LL}, it is possible for such a system to have a
phase transition provided that the interaction is sufficiently long range.
Sufficient conditions which have to be satisfied by the interaction in order for
a phase transition to exist in a one-dimensional system were given in
\cite{Dyson1}.
The papers \cite{KUH} provided the first instance of phase transition in a 
one-dimensional system, and showed explicitly that this can occur in a system 
with long-range interactions. The results
of \cite{KUH} have been extended to more general interactions and higher 
dimensional systems in Ref.~\cite{LP}.

The zero mean-reversion limit of the interest rate model $\gamma=0$ 
is the Black, Derman, Toy model in the terminal measure 
\cite{PhaseTransition}, and is equivalent 
with a lattice gas model with attractive Coulomb two-body interactions, 
placed into an external potential. This can be seen by writing the 
covariance of the Markov driver for this case as 
\begin{eqnarray}
&& -V(t_1,t_2) = \mbox{Cov}(x(t_1), x(t_2))_{\gamma=0} =
\sigma^2 \mbox{min} (t_1,t_2) \nonumber \\
&& \qquad = \frac12 \sigma^2 (|t_1 - t_2| - (t_1+t_2))\,. 
\end{eqnarray}
The first term describes an attractive linear interaction between the pair of
particles at sites $t_1,t_2$, while the
second term can be represented as their interactions with the repulsive 
external field of a static charge placed at the site $i=0$.

The one-dimensional
gas with Coulomb interaction between several types of charges was
studied, using methods very similar to those employed here, by 
Edwards and Lenard \cite{EdwardsLenard}. Our Coulomb lattice gas is
different from a usual Coulomb gas in that all particles attract each
other. The thermodynamics of a one-dimensional system with linear attractive 
potentials was considered in Ref.~\cite{Isihara}, although periodic
boundary conditions were imposed such that the resulting form of the 
interaction is different from that considered here.
A connection between stochastic processes and the
(two-dimensional) Coulomb gas was realized in a different context in 
Ref.~\cite{Dyson}.

\begin{figure}
\begin{center}
\includegraphics[height=50mm]{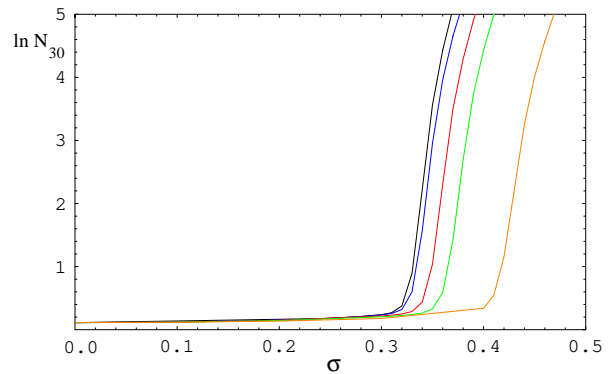}
\end{center}
\caption{(Color online) Plots of $\ln N_i(1)$ vs $\sigma$ for several values of the
mean-reversion parameter $\gamma$, with $i=30$ in a simulation with
$n=40$ quarterly time steps $\tau=0.25$. The black curve (leftmost) 
corresponds to
$\gamma=0$ and is obtained using the method used in Ref.~\cite{PhaseTransition}.
The other curves (from left to right) are obtained by explicit summation over 
the occupation numbers of the lattice gas as explained in the text:
$\gamma=0.1\%$ (blue), $1\%$ (red), $2\%$ (green), $5\%$ (orange).}
\label{fig:lnN30}
\end{figure}

The lattice gas with non-zero mean-reversion considered here differs from 
that studied by Kac and Helfand \cite{KacHelfand,Helfand} in several
respects, due to the peculiarities of the interest rate model. 

1. The presence of the $\tilde L_j$ factors requires the introduction of
single-site energies $\varepsilon_j$ associated with the lattice
sites. These energies are different and thus the space homogeneity 
of the system is lost. This space homogeneity 
was crucial for the analytical solution of the model in the thermodynamical
limit \cite{Kac,KUH,KacHelfand}. A similar approach is unlikely in this
case for this reason.

The single-site energies $\varepsilon_j$ are constrained by the
condition (\ref{tildeLi}) such that the initial yield curve $P_{0,i}$ is 
correctly reproduced.
According to this relation, $\varepsilon_j$ depends on the properties of the 
subsystem $T_{j-1}$ of the lattice gas, and must be determined by a
recursive procedure starting with the smallest subsystem $T_{n-2}$ and 
adding one lattice site at a time.

2. The two-body interaction in the lattice gas (\ref{Vdef}) contains a second
exponential term $\exp(-\gamma (t_i + t_j))$, which is not present in 
Refs.~\cite{KacHelfand,Helfand}. This is due to the fact that
the expectation values (\ref{basic}) are conditional on $x(0)=0$,
while \cite{KacHelfand,Helfand} integrate over $x(0)$.
While the new term does not have the typical form of a two-body 
interaction, its inclusion does not present any problem of principle.
Also, this term becomes vanishingly small if the subsystem
$\mathbb{T}_i$ is chosen such that $\gamma t_i \gg 1$, and the simple exponential
Kac interaction is recovered in this limit.

The equivalence of these interest rate models with lattice gases
suggests an alternative way of calibrating and simulating such models.
The expectation values $N_i(\phi)$ are usually \cite{AP,1}
computed by evaluating the 
nested integrations over the values of the Markov driver $x(t)$ at the simulation
times, using numerical approaches  such as finite difference or Monte Carlo
methods. The results (\ref{Pii}) and (\ref{Niphidef}) suggest that the
expectation values $N_i(\phi)$ can be also computed as averages over the grand 
canonical ensemble in the lattice gas. For small lattices, this can be done
by explicit summation over all possible occupation numbers ($2^n$ configurations
for a lattice with $n$ sites), while for larger lattices alternative methods 
familiar from statistical mechanics can be used, such as Gibbs sampling and the 
Metropolis-Hastings algorithm \cite{Metropolis,Hastings}. 

As an illustration of this approach, we show in Fig.~\ref{fig:lnN30} the results 
of a simulation of the BK model in the terminal measure performed by summing 
over the occupation numbers of the lattice gas. These plots show the 
multiplicative convexity adjustment $\ln N_i(1)$ for $i=30$ as function of 
$\sigma$ for several values of the mean-reversion parameter $\gamma$. The 
simulation assumed $n=40$ quarterly time steps $\tau = 0.25$, for a total 
simulation time $t_n=10$ years. The forward yield curve is flat with 
$L_i^{\rm fwd}=5\%$. The $\gamma=0$ curve is obtained using the recurrence 
method of \cite{PhaseTransition}, and the remaining curves were obtained
by computing $N_i(1)$  using (\ref{Niphidef}) by explicit summation over the
$2^{n-i-1}=512$ states of the subsystem $\mathbb{T}_{30}$ of the lattice gas. 

These results show that the transition observed in Ref.~\cite{PhaseTransition}
persists also in the model considered here. The mean-reversion $\gamma$
allows one to control the range of the two-body interaction in the lattice gas.
In the $\gamma\to 0$ limit the lattice model particles attract each other
with Coulomb potentials, while for $\gamma \neq 0$ the potential becomes
exponential and is given in Eq.~(\ref{Vdef}).
In the $\gamma\to 0$ limit the results of \cite{PhaseTransition} are recovered:
the convexity adjustment factor increases suddenly above the critical volatility
$\sigma_{\rm cr} \simeq 32\%$. 
As the mean reversion $\gamma$ is increased from zero,
the transition persists, and the critical volatility increases from its
$\gamma=0$  value. The $\gamma \to 0$ limit is well-behaved, as expected for a 
finite size lattice. 

The study of the $\gamma=0$ limit of this model presented in 
Ref.~\cite{PhaseTransition} showed that the phase transition is not visible 
under usual simulation methods used in practice for such interest rates models,
such as finite difference or Monte Carlo methods. This is due to the fact
that these methods effectively truncate the range of values of the Markovian 
driver $x(t)$ 
to a few ($\sim 5$) multiples of $\sigma \sqrt{t}$. Such a truncation
omits the contributions to the expectation values $N_i(\phi)$ which are
responsible  for the phase transition. 
The alternative method proposed here offers a possible way
to study the properties of these models, free  of these limitations.

\section{Conclusions}

We presented in this paper the exact solution of a class of interest rates 
models with log-normally distributed short rates in the terminal measure. 
The solution is formulated naturally in terms of a lattice gas with sites
corresponding to the simulation times of the model $t_i$. At each site only
one particle can be present, and the particles interact by attractive
two-body potentials $V_{ij}$ which are determined by the stochastic
process followed by the short rate. 

The analogy with the lattice gas models simplifies very much the simulation
of these models, as many of the important expectation values in the 
interest rate model can be written in closed form as averages over the
grand canonical ensemble in the corresponding lattice gas. The numerical
evaluation of these averages is straightforward for small lattices 
(few simulation times in the interest rate model), while for larger
lattices the number of configurations ($2^n$ for a lattice with $n$ sites) 
becomes too large for direct evaluation, and approximation
methods familiar from statistical 
mechanics may have to be used \cite{Metropolis,Hastings}.

We used the exact lattice gas solution to study numerically the 
Black, Karasinski model in the terminal measure with constant mean-reversion 
and volatility. This showed the appearance of a phase transition in the 
convexity adjustments of single-period interest rates, similar to that noted 
in the Black, Derman, Toy model in the terminal measure in Ref.~\cite{PhaseTransition}. 
This adds further support to the suggestion made in Ref.~\cite{PhaseTransition} 
that the presence of such a transition is generic for all interest rate models 
with log-normally distributed rates in the terminal measure. 
Although the present numerical study considered only the version of the
model with constant parameters, the method can be extended without any
major difficulty also to the more general case of time-dependent model
parameters. This is in contrast to the method of the recursion relations 
used in \cite{PhaseTransition} to solve the $\gamma=0$ limit of the model
with uniform volatility, which does not appear to be easily extended
beyond this case due to the unmanageable complexity of the resulting expressions. 

The equivalence of the interest rates models considered with 
interacting lattice gases shows that the former have a rich dynamics
which has not been fully explored. Physical intuition about the
lattice gas equivalent should give further insight into the dynamics
of the interest rate models. 
In particular, one natural question is whether a phase transition similar to 
that studied in Ref.~\cite{KUH} is present also in the lattice gas considered here,
and if it is observed also for a finite size lattice.
The analog of the van der Waals limit for this case corresponds to 
simultaneously scaling the volatility as $\sigma = \sigma_0 \gamma$ as the 
mean reversion is taken to zero $\gamma \to 0$. It would be interesting to see
if the behaviour of the system in this limit has implications also for the 
practically relevant case of non-zero volatility.

Finally, it would be interesting to investigate whether the exact solution
presented here can be extended also to other interest rate models, with 
more general distributional properties. Hopefully the lattice gas
analogy will remain useful also for more general interest rate models.

\end{document}